\documentclass[aps,superscriptaddress,floatfix]{revtex4-1}
\usepackage{bbm} 
\usepackage{amsmath} 
\usepackage{amssymb} 
\usepackage{physics} 
\usepackage[bookmarks=true]{hyperref} 
\usepackage[english]{babel} 
\usepackage[normalem]{ulem} 
\usepackage{xfrac}
\usepackage{graphicx}
\usepackage[]{xcolor} 

\allowdisplaybreaks

\begin{document}

\title{Ergotropy from Geometric Phases in a Dephasing Qubit}
\author{Fernando C. Lombardo} 
\author{Paula I. Villar}
\affiliation{Departamento de F\'\i sica, Facultad de Ciencias Exactas y Naturales, Universidad de Buenos Aires, Buenos Aires, Argentina}
\affiliation{ Instituto de F\'\i sica de Buenos Aires (IFIBA), CONICET, Universidad de Buenos Aires, Argentina}

\begin{abstract} 
We analyze the geometric phase and dynamic phase acquired by a qubit coupled to an environment through pure dephasing, establishing a direct connection between phase accumulation and ergotropy. We show that the dynamic phase depends solely on the incoherent ergotropy, reflecting its purely energetic origin. In contrast, the geometric phase exhibits a nontrivial dependence on both the coherent and incoherent contributions to the total ergotropy, encoding the interplay between coherence, dissipation, and energy extraction. By performing a perturbative expansion in the qubit–environment coupling strength, we demonstrate that, in the weak-coupling and long-time regime, the geometric phase becomes determined exclusively by the incoherent ergotropy, which coincides with the asymptotic value of the total ergotropy reached under decoherence. These results provide a clear physical distinction between dynamic and geometric phases in open quantum systems and establish geometric phases as sensitive probes of energetic resources. Furthermore,~in superconducting circuit implementations, our findings suggest that the ergotropy of a two-level system could be inferred indirectly from geometric-phase measurements using standard techniques such as quantum state tomography.

\end{abstract}

\maketitle

\section{Introduction} \label{sec:intro}

In quantum thermodynamics, ergotropy refers to the maximum amount of work that can be extracted from a quantum system through suitable quantum \mbox{operations~\cite{brandao, horo2, ergo1}}. It~serves as a measure of the system's ability to convert internal energy into usable work~\cite{binder} and plays a crucial role in optimizing quantum processes. By optimizing quantum operations, the maximum extractable work can be achieved, making ergotropy a fundamental concept in the study of quantum thermodynamics and energy management within quantum systems. Unlike classical thermodynamics, where work extraction is determined solely by the energy differences between states, ergotropy provides a broader perspective. It~not only takes energy differences into account but also considers the potential for work extraction due to quantum coherence. Quantum coherence refers to the system’s ability to maintain superposition states, which can be harnessed to extract work. Thus, ergotropy offers a more comprehensive measure of the ability of quantum systems to extract work~\cite{ergo2}. This metric accounts for both the energy differences between states and the effects of quantum coherence. Since isolating a quantum system from its environment is nearly impossible, studying work extraction from open quantum systems becomes crucial~\cite{ergo3}. In most research on work extraction from open quantum systems, the environment with which the system interacts is typically assumed to be in equilibrium and to have a significant number of degrees of freedom~\cite{erg4}. 

Recently, authors in Ref.~\cite{iran} investigate the extraction of work from quantum coherence in non-equilibrium environments, focusing on the ergotropy. The authors analyze the system's dynamics under both Markovian and non-Markovian regimes, considering the effects of non-stationary environmental interactions.  A key finding of~\cite{iran} is that work extraction is enhanced in non-equilibrium environments compared to thermal equilibrium conditions. The study distinguishes between the coherent and incoherent contributions to ergotropy, demonstrating that the coherent part increases as the environment deviates from equilibrium. This suggests that quantum coherence can be more effectively harnessed for work extraction when the system is exposed to a fluctuating or time-dependent environment.  In Ref.~\cite{Du}, 
based on the method of measuring ergotropy with an ancilla qubit, both the coherent and incoherent components of the ergotropy for the non-equilibrium state were successfully measured. The increase in ergotropy induced by the increase in the coherence of the system was observed by varying the coherence of the state.

In~\cite{Li25}, authors have experimentally resolved both coherent and incoherent ergotropy in a superconducting transmon qubit using three distinct work extraction protocols. \mbox{Their results} demonstrate that while quantum coherence does not affect the energy of a state, it fundamentally determines the amount of extractable work. By varying the Rabi angle of the initial state, they observe a clear positive correlation between coherence and coherent ergotropy, and a negative correlation with incoherent ergotropy. This suggests that the choice of initial state can be optimized based on the decoherence characteristics of the quantum battery, such as storing coherent ergotropy to prevent energy relaxation, or storing incoherent ergotropy to avoid dephasing.

These results provide valuable insights into the role of environmental dynamics in quantum thermodynamics and highlight potential applications in the design of energy-efficient quantum technologies operating in realistic, non-equilibrium conditions.  

Furthermore, due to Berry's seminal work~\cite{berry}, the geometric phase (GP) is understood as a phase accumulated by a quantum system along its evolution. Since then, it has attracted significant interest in quantum information and other branches of physics. Berry~demonstrated that quantum systems can acquire phases of geometric nature, introducing an additional phase related to the geometry of the state space during adiabatic evolution, alongside the conventional dynamical phase. Initially formulated for adiabatic and cyclic evolutions of isolated systems, this idea has been extensively generalized. \mbox{While many} of these extensions have focused on pure states, the exploration of GPs in mixed states has gained considerable attention, especially due to their potential in implementing quantum logic gates under realistic physical conditions. In this context, an alternative definition of GPs for non-degenerate density operators based on quantum interferometry was introduced~\cite{Sjoqvist}. Subsequently, a kinematic description of the mixed-state GP was provided, extending its definition to degenerate density operators~\cite{Singh}. More recently, GPs in open quantum systems have been studied, revealing that these seemingly disparate approaches are interconnected within a unified framework~\cite{Zanardi}. The impact of various decoherence mechanisms, such as dephasing and spontaneous decay, on GPs has been analyzed~\cite{Carollo}. \mbox{Additionally,~it} has been shown that geometric phases can be generated through modifications solely in the reservoir interacting with a small subsystem~\cite{Carollo2}.

GPs are fundamental in quantum computing, offering a pathway to achieve fault tolerance. However, practical implementations of quantum computing inevitably face decoherence. Therefore, appropriately generalizing the geometric phase from unitary to non-unitary evolution is essential for assessing the resilience of geometric quantum computation. This generalization has been addressed by proposing a functional representation of the GP, achieved by removing the dynamical phase from the total phase acquired by the system through a gauge transformation~\cite{Tong}.

Geometric phases have also been studied along quantum trajectories~\cite{quantum}. Research~indicates that in monitored quantum systems undergoing cyclic evolution, the accumulated geometric phase depends on the specific quantum trajectory, influenced by both unitary dynamics and environmental interactions. This results in a stochastic nature of the geometric phase due to random quantum jumps. Studies have analyzed the distribution of geometric phases in such systems, highlighting the interplay between non-adiabatic effects and environmental influences. For instance, in systems with slow driving, the phase distribution is broad because of multiple random jumps. As the driving accelerates, the number of jumps decreases, leading to a distribution centered around the no-jump trajectory, though deviations from the Berry phase occur due to non-adiabatic corrections and non-Hermitian drift terms. These findings are exemplified in studies involving a spin-1/2 particle in a time-varying magnetic field interacting with an external environment.

Geometric phases have become not only a fruitful subject for the investigation of fundamental aspects of quantum mechanics, but also an object of increasing technological interest. Owing to their nontrivial dependence on the path traced by the system’s state in the space of physical states, GPs can display high sensitivity to specific environmental or parametric variations while remaining robust against others. As a result, geometric phases can serve as highly effective resources for quantum sensing.
For instance, a velocity-dependent correction to the accumulated GP was reported for a nitrogen-vacancy center maintained at a fixed distance from a coated silicon disk mounted on a rotating platform, enabling the indirect detection of Casimir friction forces~\cite{NPJ}. The emergence of a vacuum-induced Berry phase in an artificial transmon atom was experimentally demonstrated in Ref.~\cite{Berger}, and the robustness of this geometric phase was subsequently explained in Ref.~\cite{JCLud}. Geometric phases have also been observed in superconducting circuits in which a resonator is dispersively coupled to the energy levels of a qutrit, as well as in architectures where five qubits are controllably coupled to a resonator~\cite{Song}. In the latter case, a quantum gate protocol based on this geometric phase was reported. Additional experimental observations of geometric phases can be found in Refs.~\cite{Song2,prl2010}.

{Beyond Berry’s original formulation, phase accumulation in quantum mechanics has a broader and historically rich structure, particularly in connection with the \mbox{time--energy} properties of quantum states and the evolution of wave packets. An early and insightful example is provided by Kennard~\cite{Kennard1927}, who showed that for simple types of motion, such as an electron in homogeneous electric or magnetic fields or the harmonic oscillator, the genuinely quantum-mechanical content of phase evolution can be traced back to the uncertainty relations between canonically conjugate variables and to the probabilistic interpretation of Schrödinger wave packets. From this perspective, phase accumulation emerges as an intrinsic feature of quantum wavepacket dynamics, rather than solely as a consequence of adiabatic transport in parameter space.}

{This viewpoint has gained renewed relevance in modern studies of wavepacket evolution, where the interplay between amplitude and phase encodes nontrivial physical information. In particular, recent work by Rozenman et al.~\cite{Rozenman2019} has demonstrated that wavepackets evolving in linear potentials exhibit a rich phase structure that goes beyond the standard Berry-phase framework, revealing additional phase contributions associated with the underlying time--energy structure of the dynamics. These developments provide a natural conceptual bridge between early formulations of quantum phase evolution and contemporary applications in interferometry, quantum control, and phase-\mbox{sensitive~protocols.}}

{In this broader context, the geometric phases discussed in this work can be viewed as complementing these earlier notions of phase accumulation, extending them to the regime of open quantum systems where decoherence and dissipation play a central role.
}

A natural perspective to further analyze the behavior of GPs in realistic settings is provided by energetic considerations. In open quantum systems, the presence of decoherence modifies the structure of the accumulated phase through its impact on the coherent and incoherent contributions to the system’s energy. Within this framework, the GP can be directly linked to the coherent component of the system’s ergotropy, while the dynamical phase is governed by the incoherent, population-related contribution. Decoherence~processes, such as dephasing, progressively suppress the coherent ergotropy, leading to a gradual attenuation of the geometric phase, while leaving the dynamical phase largely unaffected. This energetic decomposition offers a transparent interpretation of the robustness and fragility of geometric phases under environmental noise and highlights their intimate connection with the flow and accessibility of energy in quantum systems. 

In the following, we shall study the GP acquired by a two-level system under a dephasing evolution. We shall further find a relation between the GP and the ergotropy of the system. In the end, we shall analyze the relation between the GP acquired and the dynamic phase recently observed in dissipative quantum batteries, consisting of a minimal model of a two-level quantum system,   implemented, for example, in the lowest-energy levels of a transmon~\cite{bateries}.

\section{Geometric Phase and Ergotropy for a Dephasing Qubit}

The GP for a mixed state under non-unitary evolution is defined as follows~\cite{Tong}: 
\begin{equation}
\Phi_g = \arg \left\{\sum_k \sqrt{ \lambda_k (0) \lambda_k (\tau)} \langle \Psi_k(0)|\Psi_k(\tau)\rangle\right\}  \exp\left[{-\int_0^{T} dt \langle\Psi_k| {\dot{\Psi}_k}\rangle}\right], \label{fasegeo}
\end{equation}
where $\lambda_k(t)$ are the eigenvalues and $|\Psi_k\rangle$ are the eigenstates of the reduced density matrix $\rho_{\rm r}$ (obtained after tracing over the reservoir degrees of freedom). In this definition, $\tau$~denotes the time after which the total system would complete a cyclic evolution if it were isolated from the environment. Considering environmental effects, the system no longer undergoes a cyclic evolution. However, we consider a quasi-cyclic path ${\cal P}:t \in [0,\tau]$ with $\tau=2 \pi/\Omega$ ($\Omega$ being the system's frequency)~\cite{Tong}. Notably, the phase in Equation~(\ref{fasegeo}) is manifestly gauge-invariant, depending solely on the path in the state space, and, although defined for non-degenerate mixed states, corresponds to the unitary geometric phase in the case of a pure state (closed system)~\cite{Sjoqvist,Singh}.

It is anticipated that GPs can be observed in experiments conducted on a timescale slow enough to neglect non-adiabatic corrections but rapid enough to prevent the destruction of the interference pattern by decoherence~\cite{Gefen}. We have analyzed not only the effect of the environment on GPs and their robustness against decoherence but also the conditions under which GPs can be measured~\cite{prl2010, pra2006}. 

In this work, we aim to investigate and establish a quantitative relationship between ergotropy and the geometric phase in open quantum systems subject to dephasing. While~ergotropy quantifies the maximum amount of work that can be extracted from a quantum system, the geometric phase encodes information about the system’s evolution, particularly in the presence of dissipative and decoherent effects. With these objectives, we will introduce a simple spin–boson model and calculate the corrections to the unitary geometric phase. Furthermore, we will investigate how the geometric phase is related to the ergotropy of the system, highlighting the interplay between coherence, dissipation, and extractable work in an open quantum environment.

We will analyze how dephasing influences both quantities and explore whether there exists a fundamental link between the ability to extract work from a system and the accumulation of geometric phase. By considering different regimes of dephasing, we will examine how the coherent and incoherent contributions to ergotropy evolve and whether these contributions correlate with modifications in the geometric phase.

Our findings will provide new insights into the interplay between quantum thermodynamics and geometric effects, potentially leading to novel approaches for optimizing work extraction in noisy quantum systems. 

Hereafter, we shall study an exact model for a two-state quantum system [or quantum bit (qubit)] coupled to a thermal bath of harmonic oscillators, where decoherence is the only effect on the system-particle. The paradigmatic example is the spin--boson. In such a case, the reduced density matrix for  the two-level quantum system, the reduced density matrix ($\rho_r$) can be written as 
\vspace{-3pt}\begin{equation} \rho_r(t) = \begin{pmatrix} \rho_{11}(t) & \rho_{10}(t) \\ \rho_{01}(t) & \rho_{00}(t) \end{pmatrix}=\begin{pmatrix} \rho_{11}(0) & \rho_{10}(0)e^{-i\Omega t} e^{-\Gamma(t)} \\ \rho_{01}(0) e^{i\Omega t} e^{-\Gamma(t)} & \rho_{00}(0) \end{pmatrix}
\label{rhot}. \end{equation}

In the dephasing model, the populations remain constant during the evolution, whereas the coherences decay in a characteristic time known as {\it decoherence time} $t_D$. The~decoherence factor $\Gamma(t)$
is then defined as 
\vspace{-6pt}\begin{equation} 
\Gamma(t)=\int_0^t ~dt'~{\cal D}(t')
\end{equation}
with ${\cal D}(t)$ the diffusion term that describes the particular model (and appears in the Linblad master equation that describes the quantum evolution of the system). In the case of the spin--boson model (see~\cite{pra2006}), 
\vspace{-3pt}\begin{equation}
{\cal D}(t)=\int_0^tds\int_0^{\infty}
d\omega J(\omega) \cos(\omega s)\coth(\frac{\beta \hbar \omega}{2}),
\label{D}
\end{equation}
where $\beta=1/k_B T$ ($k_B$ is Boltzman constant) and $J(\omega)$
is the spectral density of the environment defined by the
expression $J(\omega)= \sum_n \lambda^2_n \delta(\omega-\omega_n)/2
m_n \omega_n$.

\begin{figure}
    \includegraphics[width=0.485\textwidth]{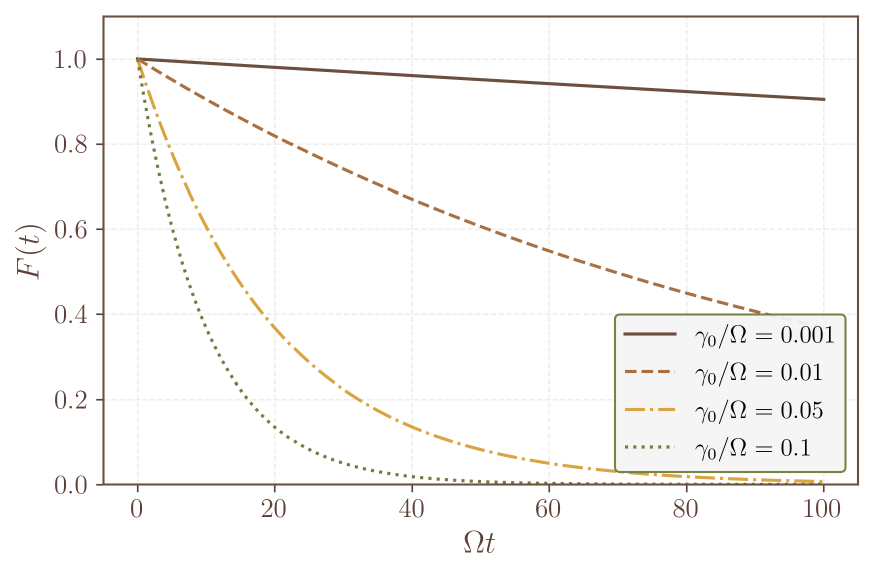}
  \hspace{0.01\textwidth} \includegraphics[width=0.487\textwidth]{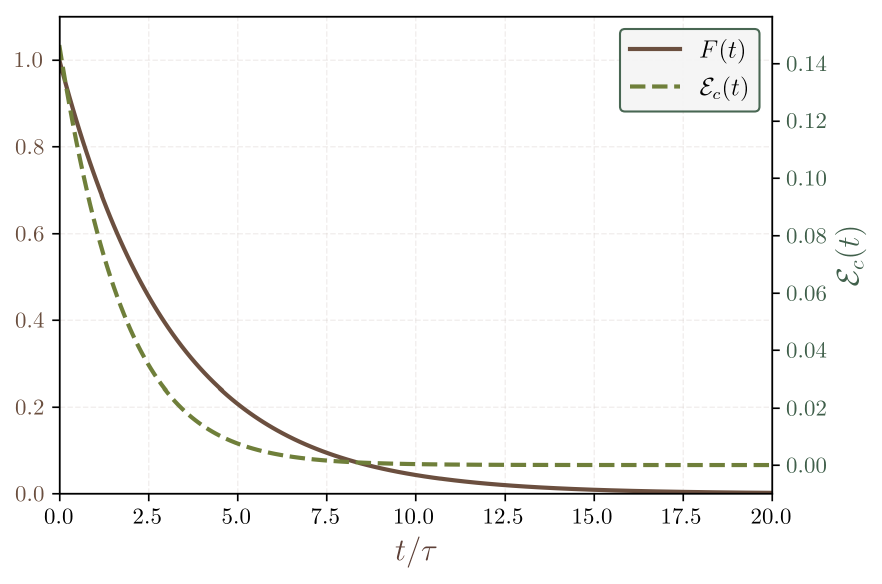}
    \caption{Dephasing factor $F(t)$ of the dephasing model. On the left, we have varied the rate  $\gamma_0/\Omega$ for a fixed  initial quantum state ($\theta=\pi/3$). On the right: we also show the coherent ergotropy ${\cal E}_c(t)$ compared to the dephasing factor for an initial angle and a fixed environment coupling ($\gamma_0/\Omega=0.05$).} 
\label{Fts}
\end{figure}

In Figure~\ref{Fts} we present the dephasing factor $F(t)=\exp\{-\Gamma(t)\}$ for the dephasing model under consideration. We can see that it is a monotonic decaying function that starts in $1$ (unitary value) and tends to zero for long times (depending on the strength of the \mbox{bath $\gamma_0/\Omega$}). On the left, we show the dephasing factor for different rates between the coupling strength of the bath and the natural frequency of the main system ($\gamma_0/\Omega$), for an initial fixed value of $\theta$, which defines the initial state of the system. On the right panel, we plot the coherent ergotropy ${\cal E}_c(t)$ as a function of time. The coherent ergotropy vanishes as dephasing effects destroy the coherences in the system.

In order to study the behavior of the geometric phase under a dephasing model, we need to compute the eigenvalues and eigenvectors of the reduced density matrix Equation~(\ref{rhot}). The eigenvalues of the density matrix of the can be written as 
\begin{equation} \lambda_{\pm}(t) = \frac{1}{2} \Bigl ( 1 \pm \sqrt{(\rho_{00}-\rho_{11})^2 + 4\rho_{01}(t)\rho_{10}(t)} ,\Bigr ). \end{equation} 
For the pure initial state, we have $\lambda_{+}(0)=1$ and $\lambda_{-}(0)=0$, and therefore, the eigenstate of interest is (see Equation~(\ref{fasegeo}))

\begin{equation} |\Psi_{+}(t)\rangle = \frac{1}{\sqrt{(\rho_{11}-\lambda_+(t))^2 + |\rho_{10}(t)|^2}} \Bigl ( (\rho_{11}-\lambda_+(t))|0\rangle - \rho_{10}(t)|1\rangle \Bigr ). \end{equation} 

Substituting into Equation~(\ref{fasegeo}), the geometric phase for a pure initial state can be exactly written as

\begin{equation} 
\Phi_g(t) = \arg \left\{ \langle \Psi_{+}(0)|\Psi_{+}(t) \rangle \right\} - \int_{0}^{T}dt \frac{{\rm Im}({\dot{\rho}_{10}(t)\rho^{*}_{10}(t)})}{(\rho_{11}-\lambda_+(t))^2 + |\rho_{10}(t)|^2}. 
\label{GP1}
\end{equation}

In order to compute the GP, we further assume an initial state, defined as
\vspace{-3pt}\begin{equation} 
|\Psi (0) \rangle=\cos(\theta/2) |0 \rangle + \sin(\theta/2) |1 \rangle .
\nonumber\end{equation}
which also defines the initial reduced density matrix as $\rho_{\rm r}(0)=|\Psi(0)\rangle \langle \Psi(0)|$.  The geometric phase is then exactly given by 
\vspace{-3pt}\begin{eqnarray} \Phi_g = - \int_{0}^{T}dt \frac{ \Omega F^2 \sin^2\theta}
{\big[\cos\theta + \sqrt{\cos^2\theta + F^2 \sin^2\theta}\big]^2 + F^2 \sin^2\theta}.  
\label{GP3}
\end{eqnarray}
 It is straightforward to prove that in the case of absence of interaction between the qubit and its environment ($F=1$), this phase reduces to the well-known Berry phase expression given by 
$\Phi_B = \pi (1 - \cos\theta)$, evaluating $T = \tau$.

\begin{figure}
    \includegraphics[width=13cm]{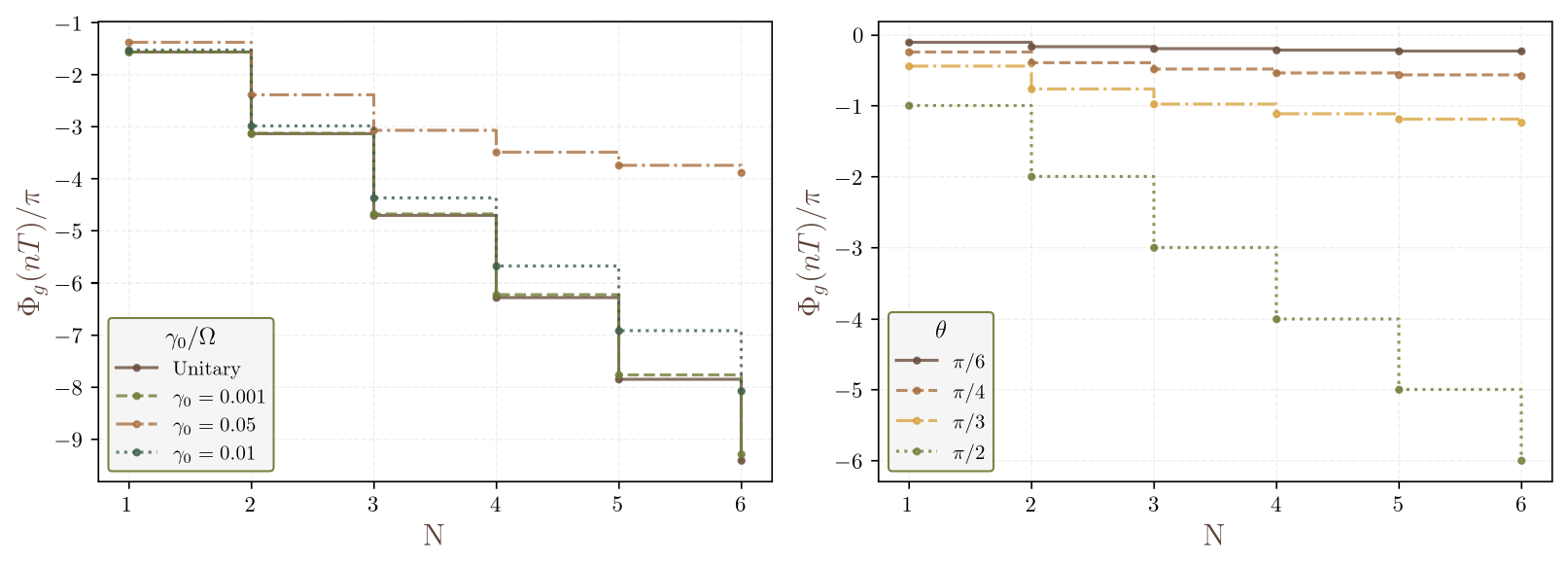}
    \caption{Geometric phase for different initial states under different dephasing environments. 
    \mbox{On the} left, we show the GP acquired for several times cycles under a dephasing model for different values of $\gamma_0$, considering  an initial state superposition state with $\theta=\pi/3$. It is easy to see that the GP remains robust since all curves are superposed.  On the right panel, we show the GP acquired by the system under a dephasing model with $\gamma_0/\Omega=0.05$ for different initial angles. }\label{escalones}
\end{figure}

In Figure~\ref{escalones}, we show the GP acquired under a dephasing evolution. On the left we present the GP as function of the number of cycles evolved for different coupling strength of the bath $\gamma_0/\Omega$ and a fixed initial angle. On the right panel we show the GP acquired for different initial states and a fixed coupling strength $\gamma_0/\Omega$.

For small angles $\theta \ll 1$, the GP is very small, while for $\theta=\pi/2$, the system acquires a GP of $\pi$ in each time cycle evolve $\tau =2\pi/\Omega$.

\subsection{Geometric Phase and Ergotropy}

Ergotropy refers to the maximum energy extractable from a quantum system through a cyclic unitary process. This cyclic unitary process can be characterized by the unitary transformation operation \( U \). The total ergotropy can be obtained as
\[
{\cal E}(\rho) = \text{Tr}(\rho H) - \text{min}_{U} \, \text{Tr}(H U \rho U^{\dagger}),
\]
where \( H \) is the Hamiltonian of the system, and the minimum is taken over the set of all unitary transformations. It has been shown that for any given arbitrary state, there exists a unique state that maximizes the above equation. This state is referred to as the passive state \( \hat{P} \). Thus, the full ergotropy can be obtained as
\[
{\cal E}(\rho) = \text{Tr} \left( \hat H \left( \hat{\rho} - \hat{P} \rho \right) \right).
\]

Considering the spectral decomposition of the density matrix $\rho_r$ and the Hamiltonian $\hat H$ as
\begin{eqnarray}
\rho_r &=& \sum_n r_n |r_n \rangle \langle r_n | \ \ \ \ \ r_1 \geq r_2 \geq \dots \geq r_n \nonumber \\
H &= & \sum_m \epsilon_m | \epsilon_m \rangle \langle \epsilon_m | \ \ \ \epsilon_1 \leq \epsilon_2 \leq \dots \leq \epsilon_m
\end{eqnarray}

So, ergotropy \( {\cal E}(\rho_r) \) is obtained as 

\[
{\cal E}(\rho_r) = \sum_{r_n, \epsilon_m} r_n \epsilon_m \left( |\langle r_n | \epsilon_m \rangle|^2 - \delta_{m,n} \right)
\]

Next, we shall examine the concept of the incoherent part of ergotropy. We begin by exploring the incoherent component of ergotropy \( {\cal E}_i(\rho_r) \). \( {\cal E}_i(\rho_r) \) can be considered as the maximum work that can be extracted from \( {\rho_r} \) without altering its coherence. Actually, \( {\cal E}_i(\rho_r) \) denotes the maximum work that can be extracted from \( {\rho_r} \) after destroying all coherence of the state using a dephasing map. \( {\cal E}_i(\rho_r) \) can be regarded as the ergotropy of the incoherent state, which has the same energy distribution as \( {\rho_r} \) with zero coherence. To calculate the incoherent ergotropy, we need to first determine the passive state corresponding to the dephased state ${\delta}_{\rho_r}$, which is denoted as \( {P}_\delta \). So, the incoherent part of the ergotropy can be obtained as
\[
{\cal E}_i(\rho_r) = {\cal E}({\delta}) = \text{Tr}\left\{ H ({\delta}_{\rho_r} -{P}_\delta )\right\} .
\]
Once the incoherent part of ergotropy is determined, the coherent part can be easily obtained as

\[
{\cal E}_c(\rho_r) = {\cal E}(\rho_r) - {\cal E}_i(\rho_r).
\]

The total ergotropy depends on quantum coherence and internal energy of the quantum system. As previously mentioned, the total ergotropy includes contributions from both the coherent and incoherent parts.
Then, for the general qubit state given by the reduced density matrix of Equation~(\ref{rhot}), we obtain 
\begin{eqnarray}
{\cal E}(\rho_r)(t) &=& \frac{1}{2} \left((\rho_{11} - \rho_{00}) + \sqrt{(\rho_{11} - \rho_{00})^2 + 4 \rho_{10}(t)\rho_{01}(t)}\right) \nonumber \\
    {\cal E}_c(\rho_r)(t) &=& \frac{1}{2} \left(- (\rho_{11} - \rho_{00}) + \sqrt{(\rho_{11} - \rho_{00})^2 + 4 \rho_{10}(t)\rho_{01}(t)}\right) \nonumber ,  
\end{eqnarray}
which implies that the incoherent part is given by
    ${\cal E}_i(\rho_r) = \rho_{11} - \rho_{00}$,
which only depends  on the populations of the quantum system that remain constant in the dephasing model. It has been proven that after decoherence sets in, the coherent ergotropy goes to zero and ${\cal E}(\rho_r) \rightarrow {\cal E}_i(\rho_r)$ for $0<\theta <\pi/2$ and $\Omega t \gg 1$. This is shown on the right side of Figure~\ref{Fts}. Therefore, we are particularly interested in the relation between ${\cal E}_c$, ${\cal E}_i$, and the non-unitary correction to the GP, induced by the environment. We call that under a non-unitary evolution the GP can be expressed as $\Phi_g = \Phi_g^u + \delta \Phi (\theta, \gamma_0, t)$, where $\delta \Phi$ represents the correction to the unitary geometric phase and encodes characteristic of the system and the environment.

In our particular dephasing model, for a general initial state as stated above, we can compute the
the incoherent ergotropy  as ${\cal E}_i(\rho_r) = \cos\theta$. The total and coherent ergotropy correspondingly are given explicitly by 
\begin{eqnarray}
{\cal E}(\rho_r)(t) &=& \frac{1}{2} \left(\cos\theta + \sqrt{\cos^2\theta + F(t)^2 \sin^2\theta}\right) \nonumber \\
   {\cal E}_c(\rho_r)(t) &=&  \frac{1}{2} \left(-\cos\theta + \sqrt{\cos^2\theta + F(t)^2 \sin^2\theta}\right)\label{ergostheta}.  
\end{eqnarray}


It is interesting to write the expression of the GP (Equation~(\ref{GP3})) in terms of the ergotropy of the system. In any dephasing model, the populations remain stable and the incoherent ergotropy remains constant  and fixed by the initial state of the system. 

For convenience, we introduce 
\begin{equation}
R(t) \equiv \sqrt{\cos^2\theta+F(t)^2\sin^2\theta},
\end{equation}
which coincides with the norm of the Bloch vector of the reduced qubit state \mbox{($R(t) = |\vec r(t)|$)}. As expected, pure dephasing leaves the longitudinal component unchanged while suppressing the transverse ones, leading to a monotonic contraction of the Bloch sphere radius. $R(t)$ measures the purity degree of the state ($R = 1$ for a pure state).
From Equation~(\ref{ergostheta}) it follows immediately that
\begin{equation}
R(t) = {\cal E}(t)+{\cal E}_c(t).
\label{eq:D_E}
\end{equation}
Using these relations, the combination appearing in the numerator of
Equation~(\ref{GP3}) can be written as
\vspace{-12pt}\begin{eqnarray}
F(t)^2\sin^2\theta
&=&
R^2-\cos^2\theta \nonumber\\
&=&
({\cal E}(t)+{\cal E}_c(t))^2-({\cal E}(t)-{\cal E}_c(t))^2 \nonumber\\
&=&
4\,{\cal E}(t)\,{\cal E}_c(t).
\label{eq:Fsin}
\end{eqnarray}
Similarly, the denominator becomes
\vspace{-6pt}\begin{eqnarray}
\big[\cos\theta+R(t)\big]^2+F(t)^2\sin^2\theta
&=&
(2{\cal E}(t))^2+4{\cal E}(t){\cal E}_c(t) \nonumber\\
&=&
4{\cal E}(t)({\cal E}(t)+{\cal E}_c(t)).
\label{eq:denominator}
\end{eqnarray}

Substituting Equations~\eqref{eq:Fsin} and \eqref{eq:denominator} into
Equation~(\ref{GP3}), the integrand simplifies drastically and the
GP can be written in a compact form
\begin{equation}
\Phi_g
= -
\int_0^T dt\;
\Omega\,\bigg(
\frac{{\cal E}_c(t)}
{2{\cal E}_c(t)+{\cal E}_i}\bigg).
\label{GP_compact}
\end{equation}
Equation~(\ref{GP_compact}) is exact and provides a reformulation of the geometric phase in terms of ergotropic quantities. In particular, the 
phase accumulation rate is proportional to the coherent ergotropy and explicitly modulated by the incoherent contribution, which enters as a static background fixed by the initial state. This expression makes transparent that coherence constitutes the essential dynamical resource underlying GP generation, while incoherent ergotropy influences the process only through its interplay with the coherent component.
In the absence of decoherence, $F=1$, the state remains pure and

\begin{equation}
{\cal E}(t)+{\cal E}_c(t)=1,
\qquad
{\cal E}_c=\frac{1}{2}(1-\cos\theta).
\end{equation}
Equation~(\ref{GP_compact}) reduces to
\vspace{-6pt}\begin{equation}
\Phi_g(T,F=1)
=
\Omega T\,{\cal E}_c= n\pi(1-\cos\theta)
\label{faseunitaria}
\end{equation}
for a cyclic evolutions with $T= n\tau$, with $n$ number of natural cycles,
recovering the standard Berry phase of a two-level system. The geometric phase is not topological in general, but acquires a solid-angle character in the coherent limit, where it becomes entirely determined by the accessible coherent ergotropy of the state. 
Equation~(\ref{GP_compact}) shows that, within a pure dephasing model (regardless of the physical characteristics of the type of environment), the geometric phase admits an exact representation entirely in terms of thermodynamic quantities, namely the coherent and incoherent contributions of the ergotropy.
In the present quantum setting, the incoherent ergotropy is fixed by the initial state through the polar angle 
$\theta$ and remains unaffected by dephasing, while the coherent ergotropy becomes a time-dependent resource that is progressively depleted by the environment.
As a consequence, the accumulation of GP is directly governed by the ratio between coherent and total ergotropy, providing a clear thermodynamic interpretation of the phase dynamics.
In the absence of decoherence, the coherent ergotropy remains constant and the geometric phase reduces to the Berry phase, thus recovering its purely geometric character and the robustness associated with unitary adiabatic evolution.
When decoherence is present, the GP is gradually suppressed as the coherent ergotropy decreases, eventually vanishing when coherences are fully lost, while the incoherent contribution provides a static background that does not  generate phase accumulation by itself.
We stress that this exact relation has been derived under the assumption of pure dephasing dynamics, and its extension to more general system–environment couplings is beyond the scope of the present work.
In the following section, we apply this formalism to evaluate the GP of a two-quantum level system interacting with an ohmic environment, considering an equilibrium bath configuration. 

\begin{figure}
    \includegraphics[width=0.9\linewidth]{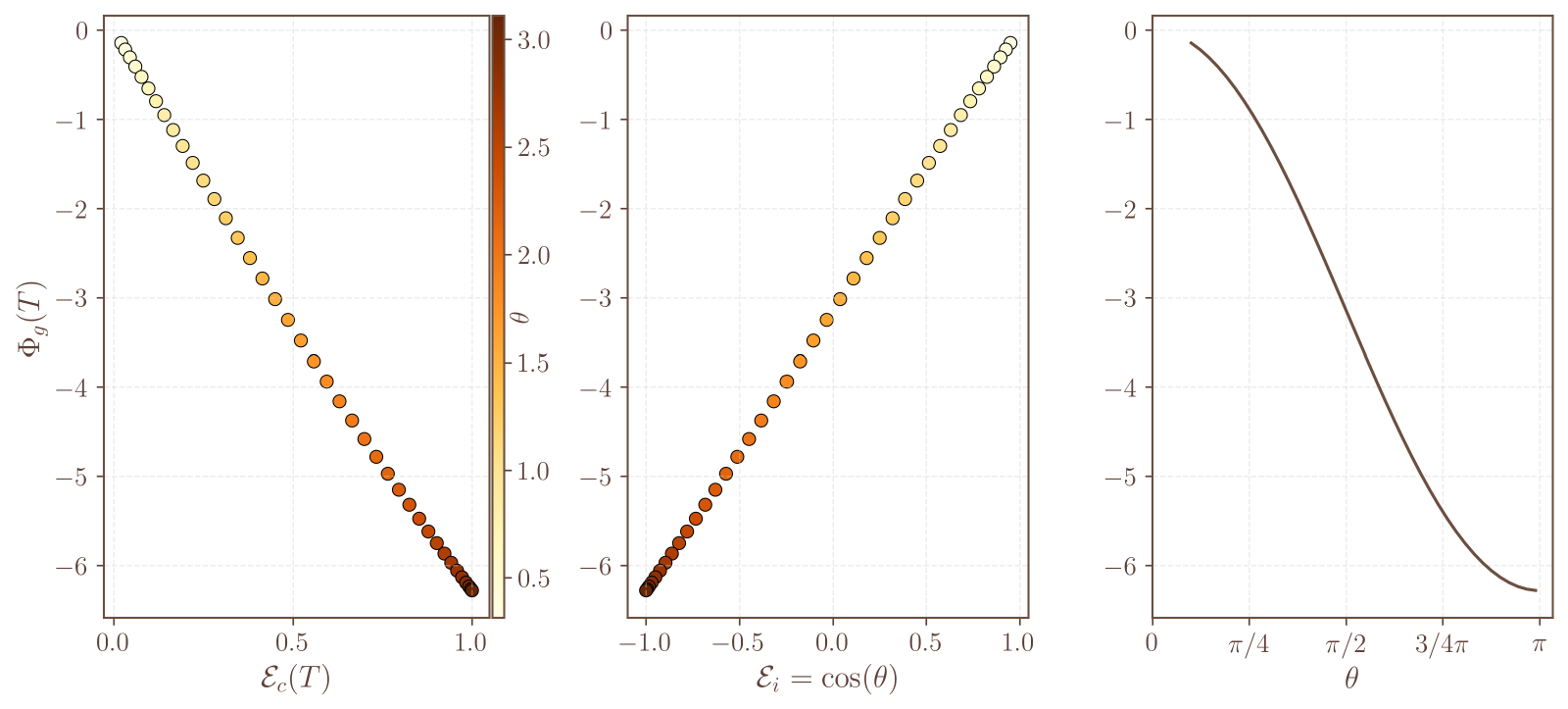}
    \caption{Geometric phase acquired in a cycle ($T=2\pi/\Omega$) showing different dependencies
    	: on the left and in the middle panels, we show the relation among the GP and the coherent ergotropy and the incoherent ergotropy, correspondingly, for different angles (lateral bar shows colors for $\theta$). \mbox{On the} right is the GP as a function of the angles for one cycle evolution. The parameters used are as follows: $\gamma_0/\Omega=0.05$, $\theta_0=\pi/3$.  }
    \label{fig:fasevsergos}
\end{figure}

Figure~\ref{fig:fasevsergos} shows, in the left and middle panels, the relationship between the GP and the coherent and incoherent ergotropy respectively. In these plots, we can conclude that the relation is quasilinear and it is almost linear for very small coupling between the system and the environment. Panel on the right shows simply the dependence of the GP with the initial Bloch angle $\theta$.

\begin{figure}
 \includegraphics[width=0.95\textwidth]{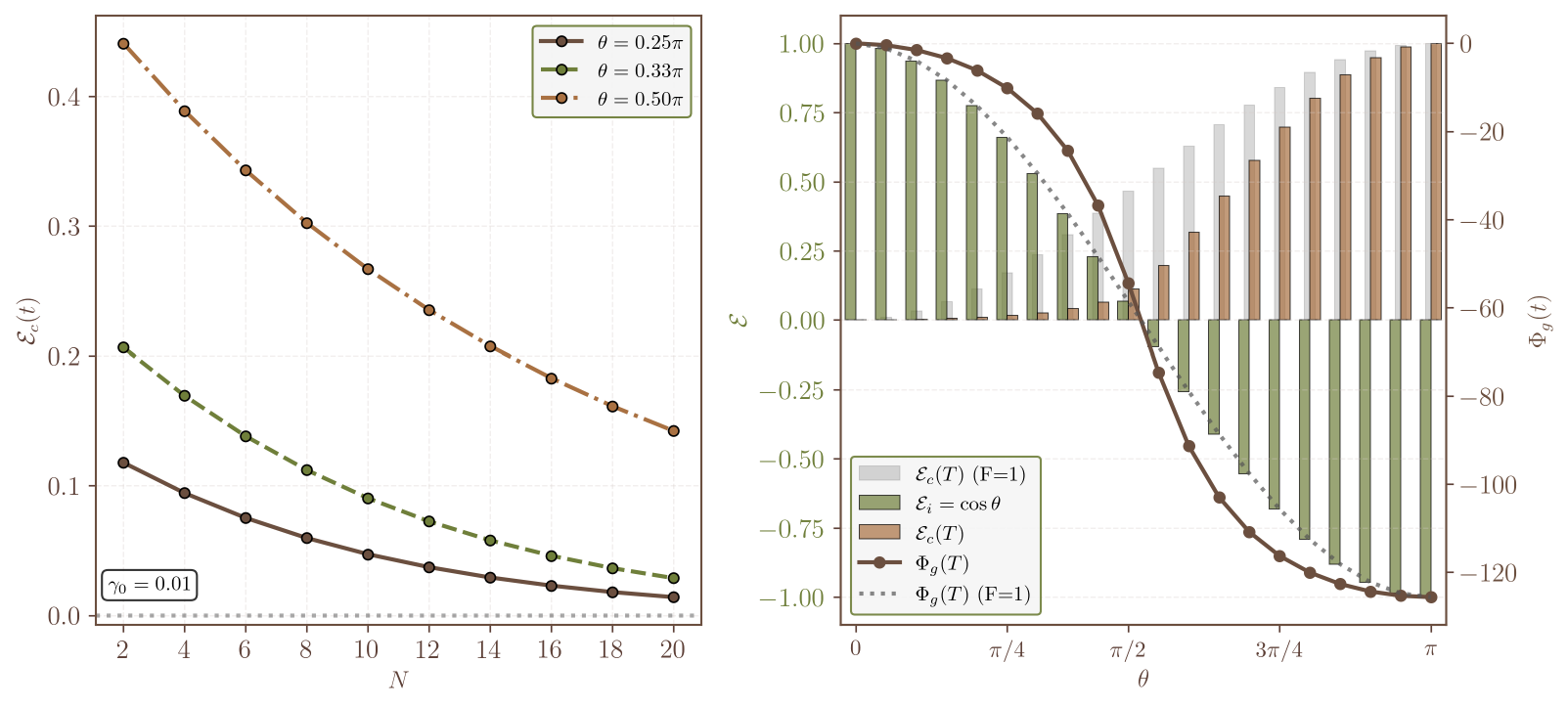}
 \caption{On the left: we present ${\cal E}_c(t)$ as function of the number of cycles elapsed for $\gamma_0/\Omega=0.01$ and different initial states. On the right: we show the coherent ergotropy (brown bars) ${\cal E}(T)$ (on the left y-axis) and  the GP acquired (on the right y-axis) after 20 cycles ($T=20\tau$). Both quantities are plotted as function of the initial state ($\theta$). ${\cal E}_c(T)=1$ for all T, since there are no coherences to destroy. The green bars represent the incoherent ergotropy (that is not modified by the environment) and gray bars are the coherent ergotropy in a unitary evolution. As a benchmark, we present the unitary GP with gray small dots.  $\gamma_0/\Omega=0.01$.}
 \label{fancy}
\end{figure}

In Figure~\ref{fancy}, on the left, we show the coherent ergotropy as a function of time for different initial angles. We can note that, since ${\cal E}_c(0)=0.5(1-\cos\theta)$, the coherent ergotropy is initially larger for angles $\theta \sim \pi/2$. In all cases, for $0<\theta<\pi/2$, ${\cal E}_c(t)\rightarrow 0$ for longer times.
On the right, we show the coherent and incoherent ergotropy and the GP as function of the initial angle $\theta$. Green bars represent the incoherent ergotropy while brown bars represent the coherent ergotropy in an evolution time $T=20\tau$. Gray bars represent the coherent ergotropy under a unitary evolution ($F=1$) as reference. The dotted line is the unitary GP acquired after 20 cycles while the brown circles represent the GP acquired in 20 cycles under dephasing. We can note the direct relation among the unitary GP and the incoherent ergotropy. We can note the strength of ${\cal E}_c$ for $\theta=\pi/2$, similarly to what happens with the GP.

\subsection{Corrections to the GP in the Dephasing Model}
  In the case of an ohmic environment at high temperature, the spectral density takes the particular form
$J(\omega)= \gamma_0/4~ \omega e^{-\omega/\Lambda}$. In that case,
$\hbar \omega << 2 k_B T$, we can
approximate ${\mathrm {coth}}(\beta \hbar \omega /2)$ in Equation~(\ref{D})
by $2 k_B T/(\hbar \omega)$ and the decoherence coefficient $\Gamma$
becomes $\Gamma= (\gamma_0 \pi k_B T) t/\hbar$, in the limit $\Lambda t \gg 1$.

The integral in Equation~(\ref{GP_compact}) can be exactly evaluated, resulting in the following 
\begin{equation}
   \Phi_g = \frac{\Omega}{2\gamma} \left(     \log[2\cos^2(\theta/2)] -  \log[\cos\theta + \sqrt{\cos^2\theta + \sin^2\theta \exp{\{-4\pi \gamma/\Omega\}}}] \right)\label{GPErgtheta}.
\end{equation}
where we have defined $\gamma \equiv \gamma_0 \pi k_B T/\hbar$, with $\gamma_0$ being the dissipative constant that includes the coupling between the system and the environment. 

Assuming the weak coupling limit in~(\ref{GPErgtheta}), we will perform a serial expansion in
terms of powers of the dissipative constant 
$\gamma_0$, and,
 consequently, the unitary phase $\Phi^{u}_g$
  is corrected
 by the presence of the environment as
\vspace{-12pt}\begin{equation}
\Phi_g = \Phi^{u}_g + \delta \Phi
\approx  \pi (1-\cos\theta) 
- \pi^2 \gamma_0 \bigg(\frac{ k_B T}{\hbar
\Omega} \bigg) \sin^2\theta \cos\theta . 
\end{equation} In the last equation we can see that the first term of the expansion is the
solution we have obtained before assuming the evolution was 
unitary, i.e., $\Phi^{u}_g=\pi (1-\cos\theta)$ (see Equation~(\ref{faseunitaria})). The second term is the GP correction rising from the interaction with a thermal ohmic bath considered herein~\cite{pra2006}.

If we were to assume the same ohmic environment but at zero
temperature, then the phase would be corrected in a significantly
different way, as one would expect since there is one timescale
lost. In that case, the factor ${\mathrm {coth}}(\beta\hbar\omega
/2)$, in the definition of $\Gamma(t)$, (Equation~(\ref{D})) can be
approximated by $1$ and the correction can be approximated as $
\delta \Phi \approx \pi^2 \gamma_0 \log(\Lambda/\Omega)
\sin^2\theta \cos \theta$, 
which is really valid for low frequency $\Omega \ll \Lambda$. 

We can rewrite these expressions in terms of the ergotropy. Therefore,  Equation~(\ref{GPErgtheta}), can be written in terms of ergotropies yields
\vspace{-6pt}\begin{equation}
   \Phi_g = \frac{\Omega}{2\gamma}   \log\left[\frac{1 + {\cal E}_i(\theta)}{ 2{\cal E}(\theta)}\right]\label{gpergexacta}.
\end{equation}
and the GP above, as a function of the ergotropy, can be expanded from Equation~(\ref{gpergexacta}):
\vspace{-12pt}\begin{eqnarray}
\Phi_g 
\approx  \Phi_g^u 
- \frac{\pi^2}{\hbar\Omega} k_B T\gamma_0 {\cal E}_i(\theta) (1 - {\cal E}_i^2(\theta)),
\end{eqnarray}
where we can see that the unitary part of the GP is linear with incoherent ergotropy, but corrections are not. In the zero temperature case, the correction can be rewritten as  
\begin{eqnarray}
\Phi_g 
\approx  \Phi_g^u 
- \pi^2 \gamma_0 \log(\Lambda/\Omega) {\cal E}_i(\theta) (1 - {\cal E}_i^2(\theta)).
\end{eqnarray}
Then, the accumulated geometric phase is proportional to the instantaneous incoherent ergotropy only in the unitary regime. Dephasing breaks this proportionality by introducing a state-dependent geometric factor. 

{It is worth emphasizing that the apparent dependence of the geometric phase solely on the incoherent ergotropy arises only under additional approximations. In particular, when expanding the geometric phase perturbatively in powers of the system–environment coupling strength and evaluating the phase after a fixed evolution time corresponding to a single cycle, the leading-order contribution becomes independent of the time-dependent coherent ergotropy. In this weak-coupling, fixed-time limit, the incoherent ergotropy, being constant under pure dephasing, sets the dominant contribution to the phase correction. Beyond this perturbative regime, however, the full expression shows that geometric phase accumulation is governed by the interplay between coherent and incoherent ergotropy, with coherence remaining the genuinely dynamical resource (see Equation~(\ref{GP_compact})).
}

\begin{figure}
 \includegraphics[width=0.49\textwidth]{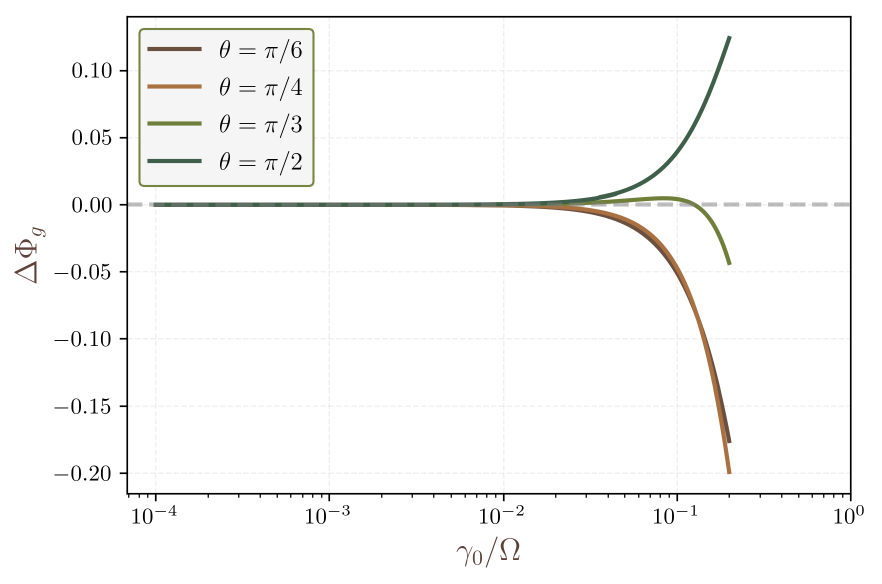}
    \hfill    \includegraphics[width=0.49\textwidth]{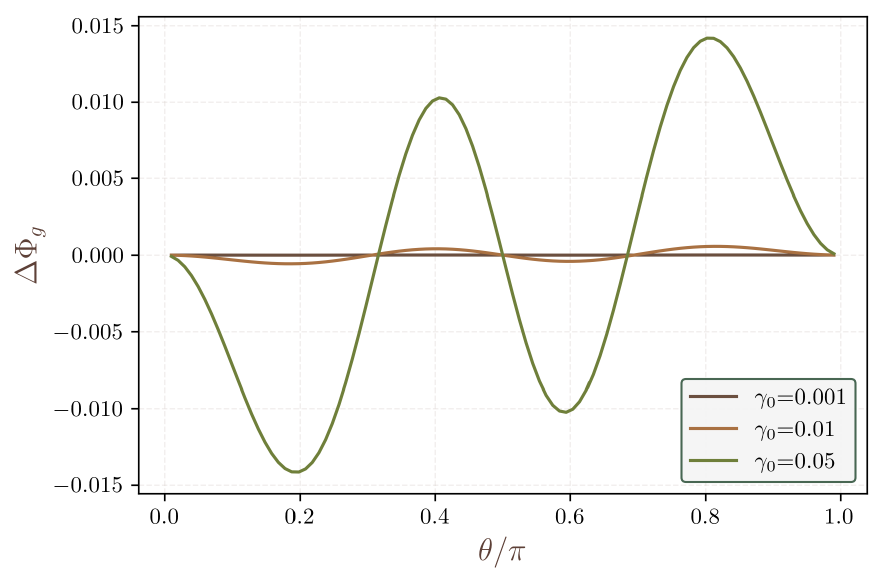}
    \caption{ $\Delta \Phi_g(T)=\Phi_{\rm exact}- \Phi_{\rm approx}$. On the left: $\Delta \Phi_g(T)$ for a zero-T dephasing model as a function of the different rate $\gamma_0/\Omega$ for a different initial angles. On the right: $\Delta \Phi_g(T)$  under a zero-T dephasing model as function of the initial angles for different $\gamma_0/\Omega$.}
    \label{Aproximadas}
\end{figure}

In Figure~\ref{Aproximadas}, the left panel shows the difference between the exact GP and the approximate expression as a function of the rate $\gamma_0/\Omega$. For weak coupling, the approximate expression reproduces the exact GP very well. For stronger couplings, the initial state plays a significant role. The right panel shows the difference as a function of the initial angle $\theta$.
We can also calculate these corrections for zero-temperature non-ohmic environments. However, the geometric structure of the corrections will always be the same, with only the coefficient preceding the correction varying, depending on the type of environment considered~\cite{pra2006}. Therefore, the decoherence time will vary with each type of environment, as will the time over which coherent ergotropy changes.


\subsection{Geometric and Dynamic Phases}

Recalling that $E(t)= {\rm Tr}[\rho(t) H] = \Omega/2 \langle \sigma_z \rangle = \Omega/2 \cos\theta$,
\vspace{-6pt}\begin{equation}
\Phi_{\rm dyn}(t)= - \int_0^T dt~E(t) \sim  -\Omega {\cal E}_i ~t \, , \end{equation}
measures the energy stored by the system. This is an extensive property, insensible to dephasing. 
We can note that the GP acts as an integral measure of accessible coherent energy along the evolution, whereas the dynamical phase counts total energy irrespective of coherence.

Looking at the rate between the different phases 
\vspace{-6pt}\begin{equation}
\frac{d\Phi_g}{d\Phi_{\rm dyn}} =\frac{2}{{\cal E}_i}
\frac{(2{\cal E}_c+ {\cal E}_i)^2-{\cal E}_i^2}
{4({\cal E}_c+{\cal E}_i)^2 + (2{\cal E}_c + {\cal E}_i)^2 - {\cal E}_i^2},
\end{equation}
we can study the particular case of small angles:
\begin{eqnarray}
{\cal E}_c(t) &\simeq& \frac{1}{2}\bigg[\frac{\theta^2}{2} + \frac{1}{2}(F(t)^2-1)\theta^2 \bigg] \sim \frac{F(t)^2}{4} \theta^2  \nonumber \\
2{\cal E}_c(t) + {\cal E}_i &\simeq& 1 + {\cal O}(\theta^3).
\end{eqnarray}
Finally, for small angles, we obtain the lineal relation
\vspace{-6pt}\begin{equation}
\Phi_g^u \approx \frac{F(t)^2 \theta^2}{2} \Phi_{\rm dyn},
\end{equation}
which shows a direct relation between the geometric phase acquired by the system and the dynamic phase. For a unitary evolution, $F(t)=1$, and the proportionality constant is fixed by the initial state. 
For a weak environmental effect, that is to say weak coupling, the lineal relation is lineal between both phases, and the slope is determined by $\gamma_0$ and $\theta$. \mbox{Finally, for} a strong dephasing environment $F(t)\rightarrow 0$ in a decoherence time $t_D$ and the geometric phase will be destroyed. However, the dynamical phase grows in time.

\begin{figure}
\includegraphics[width=0.49\textwidth]{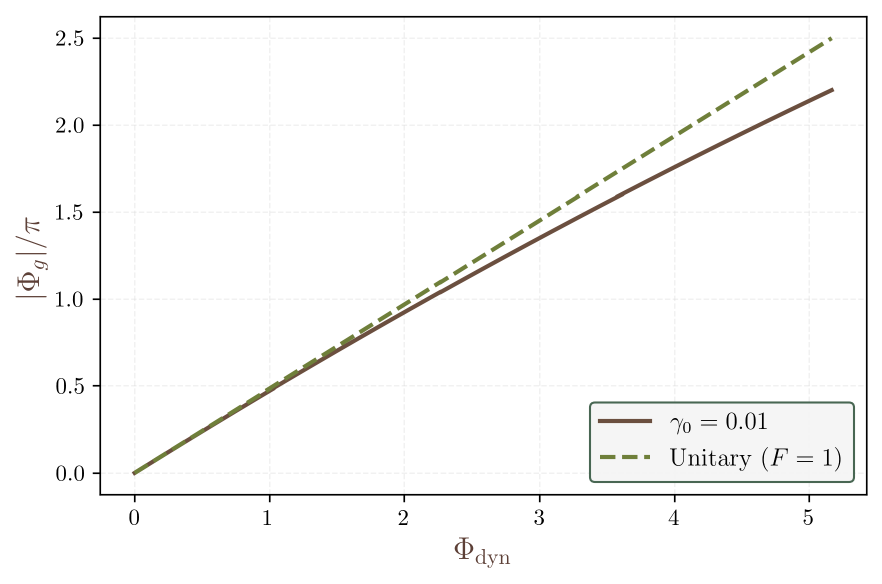}
    \hfill    \includegraphics[width=0.49\textwidth]{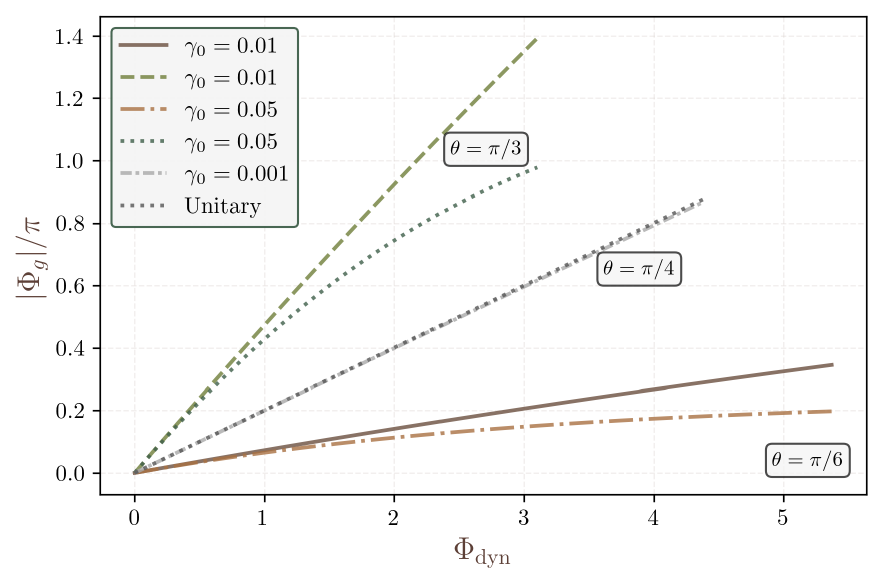}
    \caption{On the left: $|\Phi_g|$  as a function of the dynamical phase $\Phi_{\rm dyn}$ for a fixed initial small angle ($\theta_0=\pi/8$) after 5 cycles, compared to the unitary GP. On the right:  the  GP $|\Phi_g|$  as a function of the dynamical phase $\Phi_{\rm dyn}$ for different coupling rates ($\gamma_0/\Omega$) and  different initial angles.}
    \label{Aproximadas2}
\end{figure}

In Figure~\ref{Aproximadas2}, we can see that the linear relation holds longer for small angles and smaller dephasing rates $\gamma_0/\Omega$. On the left panel in Figure~\ref{Aproximadas2}, we show the relation of the GP and dynamic phase during a unitary and non-unitary evolution. On the right, we show that for small values of $\gamma_0/\Omega$ the linear relation holds for longer times. It is important to note that for transmon qubits in superconducting platforms, the dephasing coefficient has a magnitude of $\rm MHz$ while the qubit transmon has $\rm GHz$, satisfying this relation. 
{The statement that the approximately linear relation between geometric and dynamical phases persists longer for small initial polar angles and weak dephasing should be understood in a quantitative, order-of-magnitude sense. For small $\theta$, the incoherent ergotropy is close to its maximal value and the coherent contribution remains dominant during most of the evolution, delaying the onset of nonlinear corrections in the geometric phase. Similarly, weak dephasing implies 
$\gamma_0/\Omega \ll 1$, so that deviations from the unitary relation accumulate only on timescales of order $1/\gamma_0$, much longer than the qubit precession period $2\pi/\Omega$. For typical transmon parameters, where $\Omega$ lies in the GHz range while pure dephasing rates are in the MHz regime, this separation of scales ensures that the linear relation holds over many precession cycles, well within experimentally relevant times}. {\mbox{The breakdown} of the approximately linear relation $\Phi_g \sim \Phi_{dyn}$ should be understood as a crossover rather than a sharp transition. A natural dimensionless control parameter is the ratio $\gamma_0/\Omega$, which quantifies the separation of timescales between dephasing and phase accumulation, together with the initial polar angle $\theta$, which fixes the relative weight of coherent and incoherent ergotropy. Again, for $\gamma_0/\Omega \ll 1$ and small $\theta$, coherent ergotropy remains dominant over many precession cycles, and the geometric phase closely follows the dynamical one. Deviations become appreciable on timescales $t \sim 1/\gamma_0$, when the coherent ergotropy is significantly depleted and nonlinear corrections in the geometric phase accumulate. Larger initial angles accelerate this crossover by reducing the initial coherent contribution, thereby lowering the threshold at which linearity breaks down}. 

This has also been observed in the charging protocol of a dissipative two-level quantum battery~\cite{bateries}. 
Quantum batteries are proposed as devices capable of storing and releasing energy through the coherent manipulation of quantum systems, harnessing phenomena such as superposition and entanglement. 
{Geometric aspects of phase accumulation have also been extensively explored in the context of weak values. In particular, the argument of the weak value of a general observable has been shown to admit a clear geometric interpretation in terms of trajectories in projective Hilbert space, both for two-level and multilevel quantum systems~\cite{Ferraz2022}. This geometric viewpoint was later extended to open quantum systems, where the behavior of weak values under dissipative dynamics and non-unitary evolution was analyzed, highlighting the interplay between geometry, decoherence, and measurement backaction~\cite{Ferraz2024}.} 

{Within the two-state vector formalism, weak values have also been investigated in scattering and reflectivity scenarios, where their phase and amplitude were shown to possess direct experimental relevance and predictive power~\cite{Dreismann2019,Dreismann2021}. While these works establish the geometric and operational significance of weak-value phases, they primarily address informational and measurement-related aspects. By contrast, in the present work the geometric phase acquired by the system itself is shown to encode the depletion of thermodynamic resources, specifically, the coherent contribution to ergotropy, under open-system dynamics, thereby providing a direct link between geometric phases and quantum battery performance.}

However, a deeper understanding of their behavior requires characterizing not only the energy levels involved but also the geometric and topological structure of the state space~\cite{LuEntropy}. In this context, the study of the geometric phase provides information that cannot be obtained from the energy spectrum alone: while energy describes the local and instantaneous properties of the system, the geometric phase depends on the path traced in parameter space and reveals the global connectivity between quantum states. In particular, when the system possesses a topological structure, the geometric phase allows one to identify robust invariants that characterize classes of states protected against perturbations. Thus, geometric and topological analysis offers a perspective complementary to the energetic one, providing a deeper understanding of the quantum mechanisms that may influence the stability, coherence, and collective properties of quantum batteries. This~approach has been used before. For example, the use of the geometric phase has been proposed to detect traces of quantum forces in a NV center in front of a dielectric surface in~\cite{NPJ}. In addition, it has been proposed to detect gravitational waves in~\cite{BerryGravitational}.

\section{Conclusions}

In this work, we have explored the interplay between geometric phases and ergotropy in open quantum systems, highlighting the fundamental role of incoherent contributions in determining the asymptotic work extractable from a quantum system. Our results demonstrate that while ergotropy encompasses both coherent and incoherent components, the geometric phase is solely sensitive to the incoherent ergotropy, which emerges in the long-time limit under the influence of decoherence and dissipation.

The geometric phase can be thought of as a resource tool to characterize the quantum system, that is to say, it is related to the quantum nature of the system. As long it is not zero, the system preserves its quantum coherences. Since the efficiency of a quantum battery is fundamentally bounded by its ergotropy, i.e., the extractable work via unitary operations, any observable that reliably encodes this quantity becomes a powerful diagnostic \mbox{tool~\cite{ergo5,bateries}}. The total ergotropy depends on both the quantum coherences and the internal energy of the system, thus combining features of the geometric phase and the energy. It can be shown that, in the long-time limit, the geometric phase accumulated by the system is determined solely by the incoherent component of the ergotropy (arising from quantum coherences), which remains robust under dephasing. This connection opens a new route for non-invasive characterization of quantum batteries, especially in platforms like superconducting circuits where full state reconstruction is feasible.

This finding establishes a novel and experimentally relevant connection between geometric phases, traditionally considered purely geometric and global in nature, and a key thermodynamic quantity such as ergotropy. In particular, it suggests a practical route for estimating the ergotropy of a quantum battery by measuring the geometric phase, especially in platforms such as superconducting circuits where quantum state tomography is routinely available.

{Finally, it is worth emphasizing that the exact relation between geometric phase and ergotropy derived in this work relies on the assumption of pure dephasing dynamics, for which energy exchange with the environment is absent and the incoherent ergotropy remains fixed by the initial state. In the presence of dissipative system–environment couplings involving energy exchange, one may expect a qualitatively richer scenario, where both coherent and incoherent contributions to ergotropy become time dependent. In such cases, geometric phase accumulation would likely reflect not only the depletion of coherence but also genuine energetic relaxation processes, potentially blurring the separation between geometric and thermodynamic contributions. Exploring whether a similarly compact relation between geometric phase and thermodynamic resources can be established for dissipative models remains an open question and is left for future work.}

\end{document}